\documentclass[preprint,prl,amsmath,amssymb,showpacs]{revtex4}

\usepackage{graphicx}% Include figure files
\usepackage{dcolumn}% Align table columns on decimal point
\usepackage{bm}% bold math
\usepackage{overpic}

\begin{document}

%\preprint{}

\title{Conceal an entrance by means of superscatterer}
\author{Xudong Luo}\thanks{To whom correspondence should be
addressed. \\ Email address: luoxd@sjtu.edu.cn}
\author{Tao Yang}
\author{Yongwei Gu}
\author{Hongru Ma}

\affiliation{Institute of Theoretical Physics, Shanghai Jiao Tong
University, Shanghai 200240, People's Republic of China}

\date{\today}

\begin{abstract}
By using the novel property of the rectangular superscatterer, we
propose a design which can conceal an entrance from
electromagnetic wave detection. Such a superscatterer is realized
by coating a negative index material shell on a perfect electrical
conductor rectangle cylinder. The results are numerically
confirmed by full-wave simulations both in the far-field and
near-field.

\end{abstract}

\pacs{41.20.Jb, 42.25.Gy}

\maketitle

Electromagnetic (EM) metamaterials \cite{Pendry96,Pendry99} are
artificially engineered materials with subwavelength composites,
their effective parameters of permittivity and permeability are
desirable and can attain more wider ranges than those in natural
materials. The progress in metamaterial fabrication provides more
freedom for designing striking devices, such as perfect lens
\cite{Pendry2000,Pendry2003}, cloaks
\cite{Greenleaf,Alu,Leonhardt,Pendry2006,Schurig,Cai}, hyperlens
\cite{Jacob,Salandrino,Kildishev,Zhangx} and other invisibility
devices \cite{Chenhy,concentrator,Zhang}. These devices seem
unphysical or unattainable before, because it needs very special
permittivity and permeability tensors, for example, the negative
index material (NIM) \cite{Veselago,Pendry2000}.

More recently, a new device called superscatterer \cite{Yang} was
proposed by means of the concept of complementary media
\cite{Pendry_Rama}. In EM wave detection, it looks like a
scatterer bigger than the physical size of the device. Since the
outer region in the pair of complementary media is just the vacuum
or host medium, it might be regarded as a kind of building blocks
used to construct more complex objects, for example, a penetrable
barrier or a closed two dimensional (2D) box, which are opaque by
EM waves but penetrable by small particles. However, the building
blocks need to be nicely fitted together to obtain the effect so
that the circular shaped blocks may be inapplicable. In this
letter, we design a 2D rectangular cylindrical superscatterer and
demonstrate its interesting applications in concealing entrances.
% based on its combinability in shape.

\begin{figure}
\includegraphics{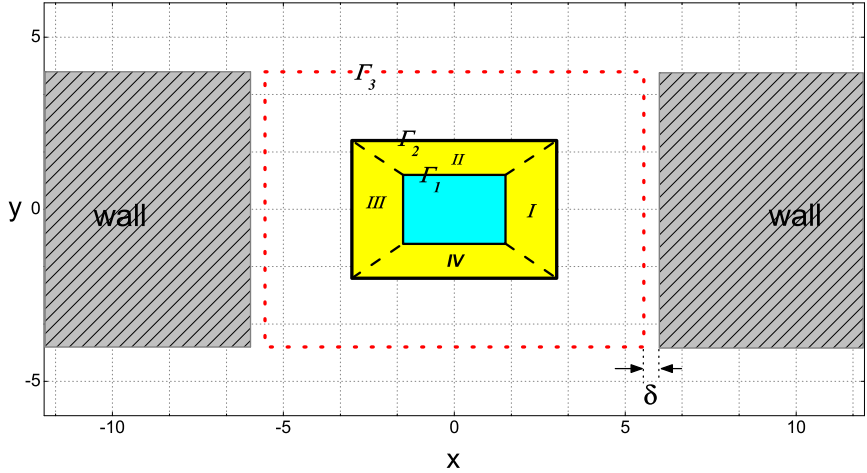}
\caption{\label{wall} The schematic demonstration of how to design
an electromagnetic wave concealed entrance by rectangular
superscatterer with a pair of complementary media. The dashed
rectangular is the domain to be concealed from electromagnetic
detection, the smaller rectangular within solid line is the
physical size of the superscatterer.}
\end{figure}

For simplicity, let's consider a perfect electrical conductor
(PEC) wall ($|y|\leq b_3$) in vacuum, any incident EM waves will
be reflected from it. If we remove the material in $|x|\leq
a_3+\delta$ from the wall, an entrance appears, and it is
inescapably detectable by analyzing the scattering EM waves.
However, superscatterer provides a possibility that the entrance
is kept but can not be detected by EM waves. Fig. \ref{wall} is
the schematic demonstration and the total procedure is displayed
as follows. Firstly, if we only remove the PEC materials in $a_3
\leq |x|\leq a_3+\delta$, it still looks like a wall when the
non-negative parameter $\delta$ tends to zero. Secondly, we
compress the big rectangular cylinder, $|x|\leq a_3$ and $|y|\leq
b_3$, into a small rectangular cylinder, $|x|\leq a_1$ and
$|y|\leq b_1$. Now, the surfaces of these rectangular cylinders
are denoted as $\Gamma_3$ and $\Gamma_1$, respectively. The
material parameters of the compressed cylinder are obtained by the
corresponding coordinate transformation. But in this case, it is a
PEC cylinder that we compressed, so the PEC boundary condition is
also kept in the surface $\Gamma_1$. Finally, in order to conceal
the entrance in EM detection, the gap between $\Gamma_1$ and
$\Gamma_3$ should be filled by a pair of complementary media: One
is the vacuum between $\Gamma_3$ and $\Gamma_2$, the other is the
NIM shell between $\Gamma_2$ and $\Gamma_1$. Here $\Gamma_2$ is
the outer rectangular surface of the NIM shell with dimensions $2
a_2$ and $2 b_2$.

This device is a superscatterer and scatters the same fields as
the uncompressed rectangular cylinder, so that the entrance
between $\Gamma_2$ and $\Gamma_3$ is concealed in EM detection.
The effective rectangular cylinder with surface $\Gamma_3$ is also
called a virtual rectangular cylinder. Although the surface of
$\Gamma_2$ can be any shape bounded between $\Gamma_1$ and
$\Gamma_3$, here we only consider a simple shape: In the $x$-$y$
plane, $\Gamma_1$, $\Gamma_2$ and $\Gamma_3$ are rectangles with
the same center and diagonals. In addition, the magnification
factor $\eta$ is defined as $a_3/a_2$ or $b_3/b_2$, which is the ratio
of the size of the virtual cylinder to the real size of this
device.

According to the discussion in  \cite{Yang}, the parameters of
complementary media can be obtained by the approach of the
transformation optics\cite{Pendry2006}. Here, we consider a simple
continuous map between the pair of complementary media: The region
of vacuum is mapped to the region of NIM shell, especially, the
surface $\Gamma_3$ is mapped to $\Gamma_1$ but the surface
$\Gamma_2$ is mapped to itself. For the 2D rectangular cylindrical
superscatterer, the NIM shell is separated to $4$ regions as shown
in Fig. \ref{wall}, and the coordinate transformation equations
for region $I$ are
\begin{equation}\label{eq1}
\begin{array}{rcl}
x^\prime &= & \displaystyle -\frac{a_2-a_1}{a_3-a_2}\,
x+\frac{a_3-a_1}{a_3-a_2}\, a_2, \\
y^\prime &= & \displaystyle -\frac{a_2-a_1}{a_3-a_2}y
+a_2 \frac{a_3-a_1}{a_3-a_2} \, \frac{y}{x},\\
z^\prime & =&z.
\end{array}
\end{equation}
The relations of $a_1 < a_2 < a_3$ provide a folded geometry
\cite{Leonhardt2006,Milton}, so the NIM appears. In addition, it
becomes the usual rectangular concentrator if the relations are
$a_1 < a_3 < a_2$, or it describes a rectangular cloak if $a_1 <
a_2$ and $a_3=0$. For comparison, the similar transformation
equations for square cloak can be found in paper
\cite{concentrator}.

In the new coordinate system, using the Jacobian transformation
matrix
\begin{equation}\label{eq2}
\Lambda^{i^\prime}_{i}=\frac{\partial
x^{i^\prime}}{\partial x^i},
\end{equation}
the components of relative permittivity tensor
$\overline{\overline{\epsilon^\prime}}$ and relative permeability
tensor $\overline{\overline{\mu^\prime}}$ \cite{Schurig2} can be
put in the following form,
\begin{equation}\label{eq3}
\begin{array}{rcl}
\epsilon^{i^\prime j^\prime} &=&
[\det \Lambda]^{-1} \Lambda^{i^\prime}_i \Lambda^{j^\prime}_j \delta^{ij}, \\
\mu^{i^\prime j^\prime} &=& [\det \Lambda]^{-1}
\Lambda^{i^\prime}_i \Lambda^{j^\prime}_j \delta^{ij},
\end{array}
\end{equation}
where $\delta^{ij}$ is equal to $1$ for $i=j$ and equal to $0$
otherwise. By straight forward calculations, the relative
permittivity and permeability tensors in region $I$ are obtained
as follows,
\begin{equation}\label{eq4}
\overline{\overline{\epsilon^\prime}}
=\overline{\overline{\mu^\prime}} =\left(\begin{array}{ccc}
 \frac{\Delta_1}{(a_3 -a_2)^2 \,x^{\prime}} & \Delta_2  & 0 \\
   \Delta_2 &
  \frac{(a_3-a_2)^2\,{x^{\prime}}^4
  +(a_3-a_1)^2\,{a_2}^2\,{y^{\prime}}^2}{{x^{\prime}}^3\,\Delta_1 } & 0 \\
   0 & 0 & \frac{\Delta_1 }{(a_2 -a_1)^2\,x^{\prime}}
\end{array} \right),
\end{equation}
where
\begin{equation}\label{eq5}
\begin{array}{rcl}
\Delta_1 &=& (a_3-a_2)^2 x^{\prime}-(a_3-a_2)(a_3 -a_1)\,a_2, \\
\Delta_2 &=& \displaystyle -\frac{(a_3-a_1) \,a_2\,y^{\prime}}{(
a_3 - a_2) \,{x^{\prime}}^2}.
\end{array}
\end{equation}
Due to the symmetry of this device, the material parameters
$\epsilon^{i^\prime j^\prime}$ and $\mu^{i^\prime j^\prime}$ in
the region $III$ can be readily obtained by rotating the tensors
in \eqref{eq4} by $\pi$ around the $z$-axis. But for the regions
$II$ and $IV$, besides of rotating the tensors in \eqref{eq4} by
angle $\pi/2$ and $3 \pi/2$, we need replace all $a_k$ with $b_k$,
$k=1,2,3$. Here, it should be remarked that the obtained material
parameters are continuous at the interfaces among these regions.
Subsequently, the materials interpretation \cite{Schurig2} means
the tensors in \eqref{eq4} and its counterparts in other regions
just describe the material parameters of rectangular
superscatterer, if we substitute the primed indices by unprimed
indices.

In order to demonstrate the performance of the designed device in
Fig. \ref{wall}, full-wave simulations are performed by finite
element solver of the Comsol Multiphysics software package. Here
the device parameters $\{ a_1,a_2,a_3\}$ and $\{ b_1,b_2,b_3\}$
are $\{0.03\hbox{m}, 0.06\hbox{m},0.12\hbox{m} \}$ and
$\{0.015\hbox{m}, 0.03\hbox{m}, 0.06\hbox{m}\}$, respectively, so
that the magnification factor $\eta$ is equal to $2$. In this
case, we consider 2D transverse electric (TE) polarization
incident waves (whose electric field is along $z$-axis) with
frequency $7$ GHz, it means the distances between device and
walls, or the width of entrance, are about $1.4 \lambda$ where
$\lambda$ is the wavelength.%
%\textbf{In order to fabricate such a
%material, it is desirable that tensors
%$\overline{\overline\epsilon}$ and $\overline{\overline\mu}$ are
%diagonal by presented in their eigenbasis. However, like the case
%of square cloak \cite{concentrator}, it needs different rotation
%angle at different position in region $I$.}
The ranges of material parameters along their local principle axis
are taken as follows: $\mu_x, \mu_y \in [-20.5,-0.05]$, and
$\epsilon_z \in [-8,-2]$, which are finite and in a range of
possible fabrication. A tiny absorptive imaginary part ($\sim
10^{-12}$) is added to $\overline{\overline\epsilon}$ and
$\overline{\overline\mu}$ due to the inevitable losses of the NIM.

\begin{figure}
\includegraphics{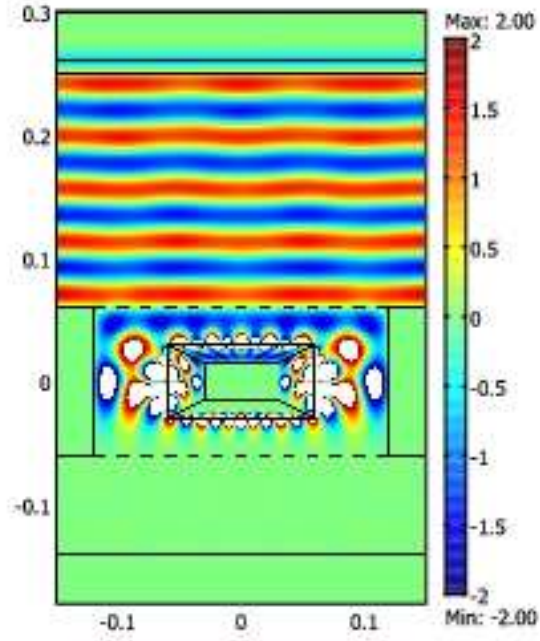}
\caption{\label{periodic} Snapshot of the total $E_z$ field around
the superscatterer concealed entrance. The electric field was
totally reflected so that the electric field on the  other side of
the entrance is barely zero.}
\end{figure}

We first simulate the case of plane wave incidence with unit
amplitude. It is interesting to calculate the scattering
properties of a penetrable barrier, which is a layer of 2D
photonic crystal and the primitive cell is the superscatterer and
a rectangular PEC cylinder. In the $x$-$y$ plane, the scale of the
virtual cylinder is $0.24$m $\times$ $0.12$m and those of PEC
cylinder is $0.06$m $\times$ $0.12$m. The primitive cell is shown
in Fig. \ref{periodic}, and the pattern of electric field is
calculated by the finite element solver with periodic boundary
condition at $x=\pm 0.15$m. One can find the superscatterers do
work as building blocks and the layer of 2D photonic crystals
looks like a mirror (a PEC wall) although it is penetrable for
small particles. Here, the bounds of the amplitude of electric
field in Fig. \ref{periodic} set from $-2$ to $2$ for clarity. The
white flecks show the regions where the value of fields exceed the
bounds. It appears in the complementary media and the highest
value of the field in the flecks is about $10^1$, which comes from
the dominative high-$m$ modes as in the circular version
\cite{Yang}.

\begin{figure}
\begin{overpic}[scale=0.6,bb=0 0 352 264]{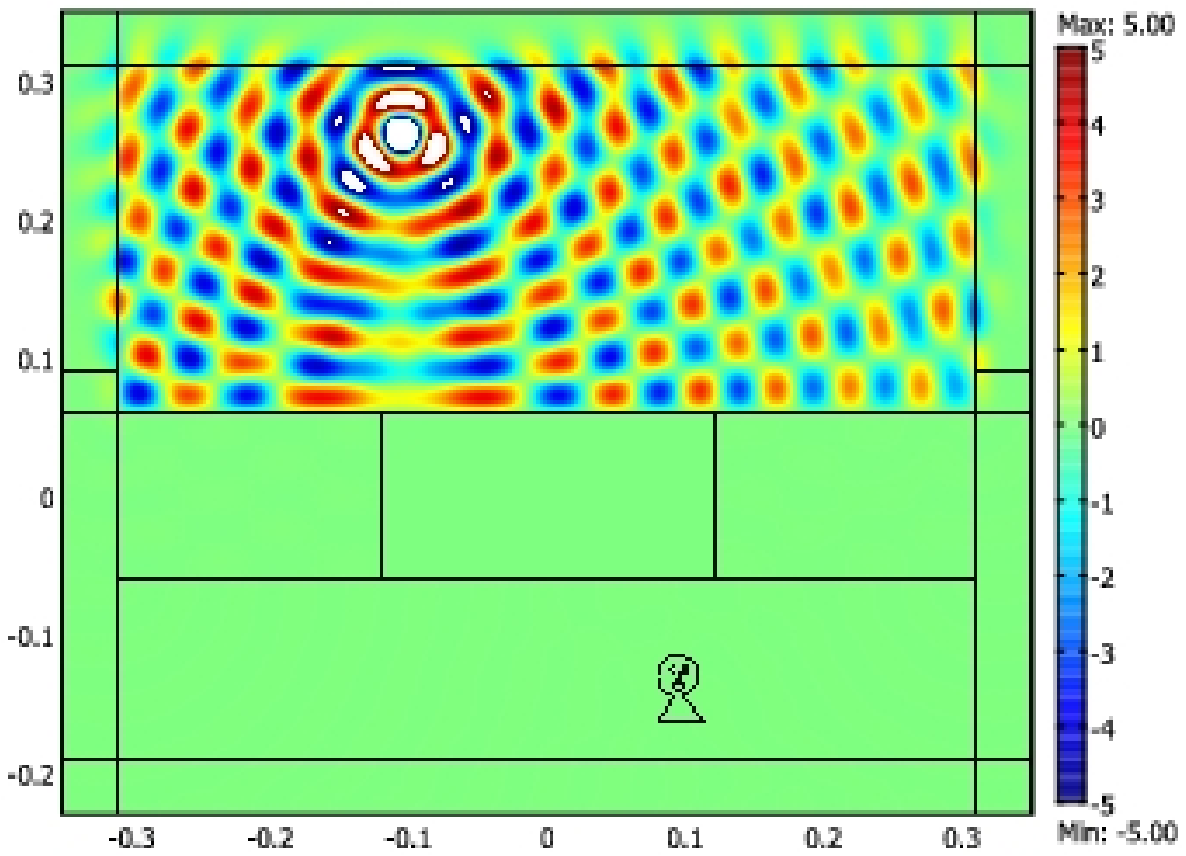}
\put(72,14){\bf(a)}
\end{overpic}
\begin{overpic}[scale=0.6,bb=0 0 352 264]{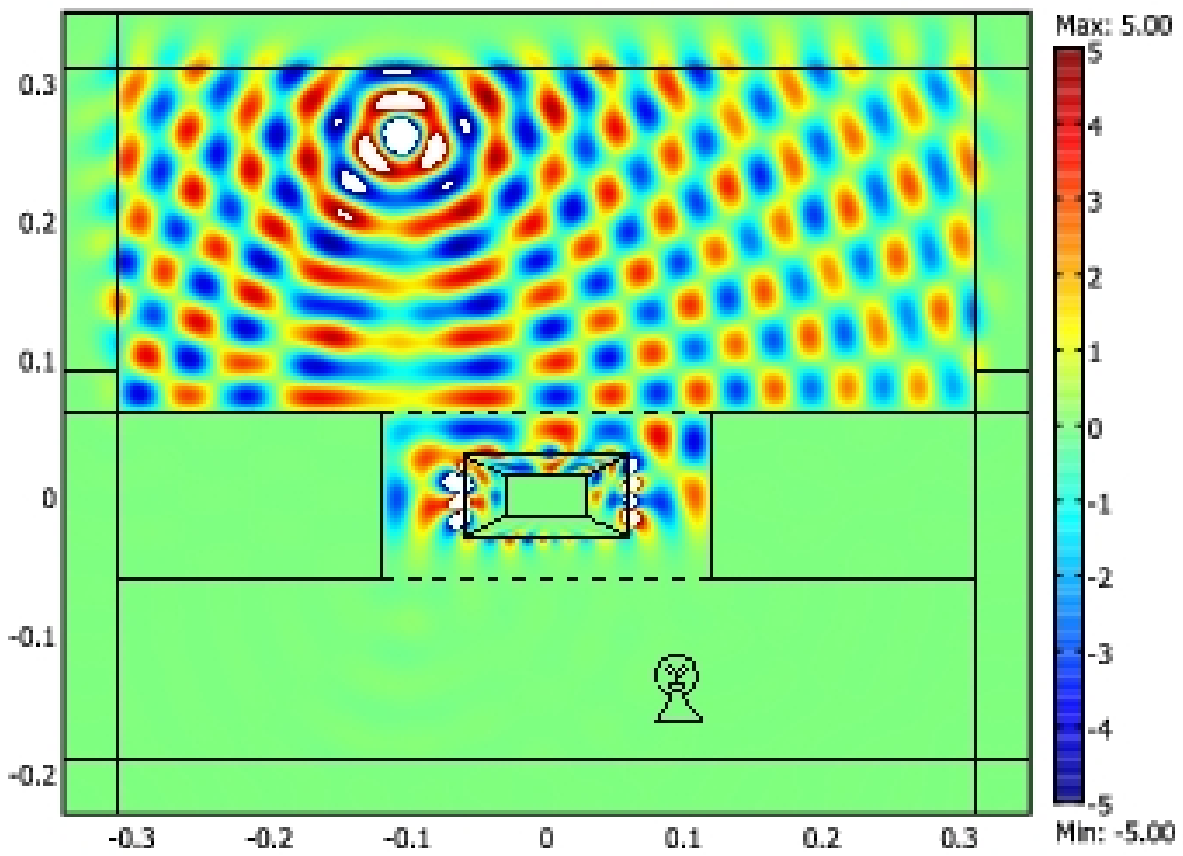}
\put(72,14){\bf(b)}
\end{overpic}
\begin{overpic}[scale=0.6,bb=0 0 352 264]{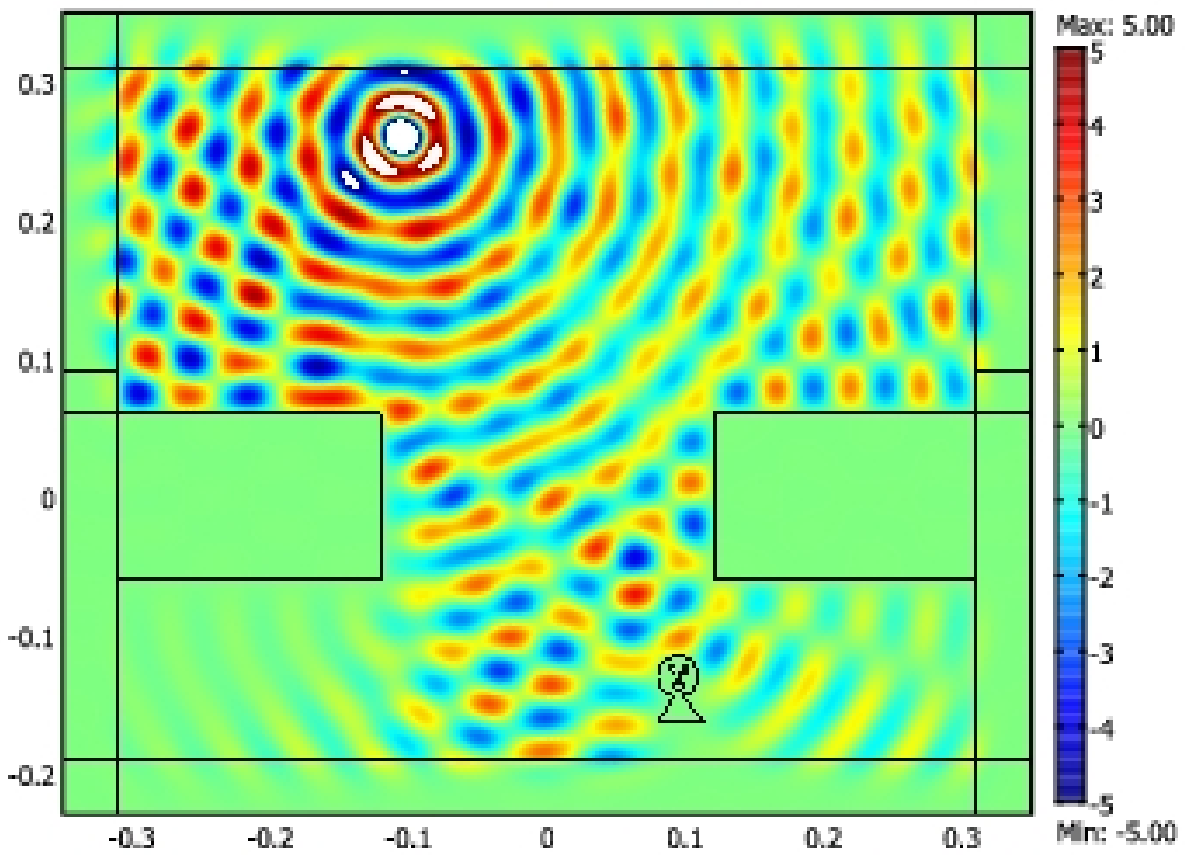}
\put(72,14){\bf(c)}
\end{overpic}
\begin{overpic}[scale=0.6,bb=0 0 352 264]{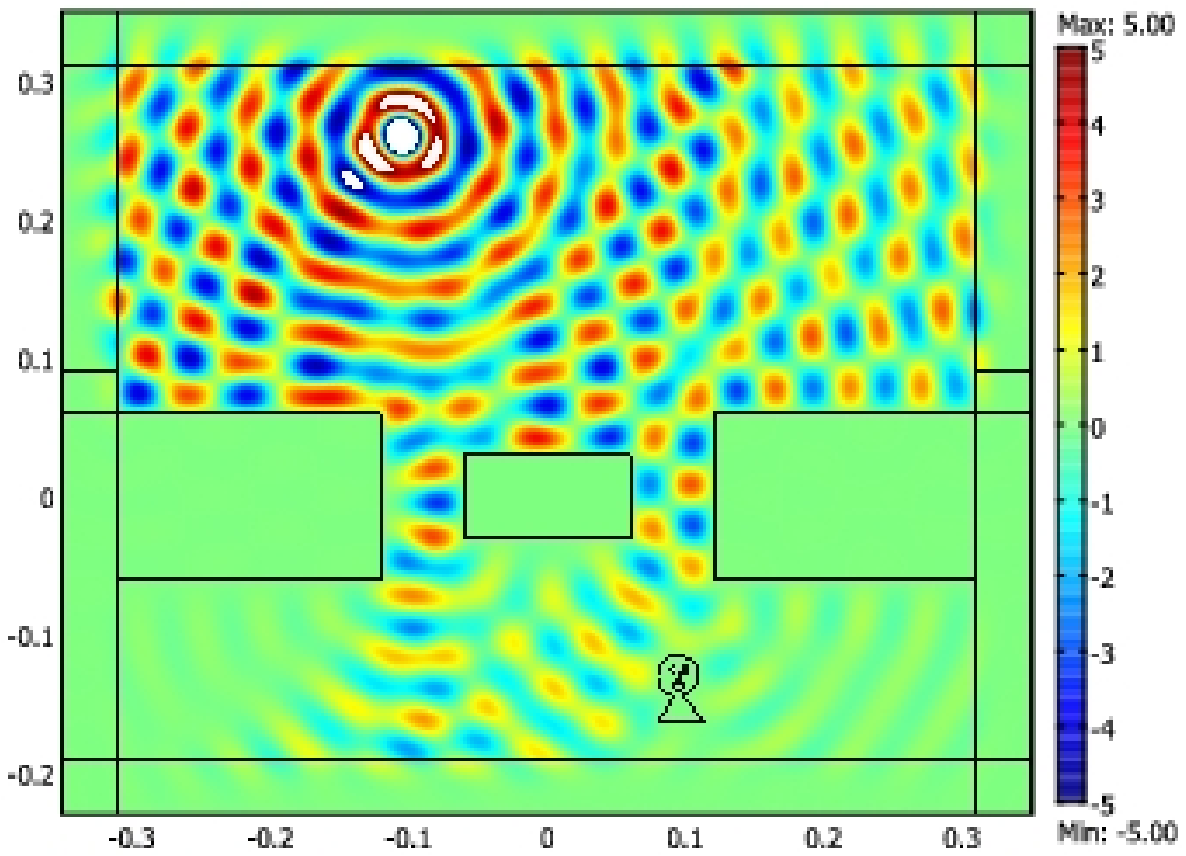}
\put(72,14){\bf(d)}
\end{overpic}
\caption{\label{point} Snapshot of the distribution of the total
electric field with a line source. (a)  The total electric field
distribution when a PEC wall is present.  (b)  The total electric
field distribution when an entrance is concealed by the
superscatterer.   (c) - (d) The total electric fields with an
entrance unblocked and blocked by a rectangular PEC cylinder with
the same physical size of the superscatterer, respectively.}
\end{figure}

We also investigated the near-field performance of the device by
cylindrical wave incidence. In this case, there is a concealed
entrance in the wall, and the electric line source carries a
current of $0.001$A in $z$-axis and located at $(-0.105\hbox{m},
0.26\hbox{m})$. The surrounding regions are perfect match layer
(PML) regions, and an object with PEC boundary is placed under the
wall. Fig. \ref{point}(a) is the snapshot of total electric field
induced by a PEC wall, and Fig. \ref{point}(b) demonstrates how
the entrance is concealed by a rectangular superscatterer in EM
detection. For comparison, we have also simulated the results of
removing the device (See in Fig. \ref{point}(c)), or replacing the
device with a rectangular cylinder with PEC boundary, which has
the same dimensions as the device (See in Fig. \ref{point}(d)).
There have some white flecks, in which the value of fields exceeds
the bounds $[-5 \hbox{V/m}, 5 \hbox{V/m}]$, around electric line
source, because the field trending to source will increase
quickly. Other white flecks appears in the complementary media and
the highest value of the field in the flecks is about $10^1$ V/m.
Comparing the patterns of electric field in Fig. \ref{point}, one
can find the entrance is indeed concealed in EM detection.

\begin{figure}
\begin{overpic}[scale=0.6,bb=0 0 352 264]{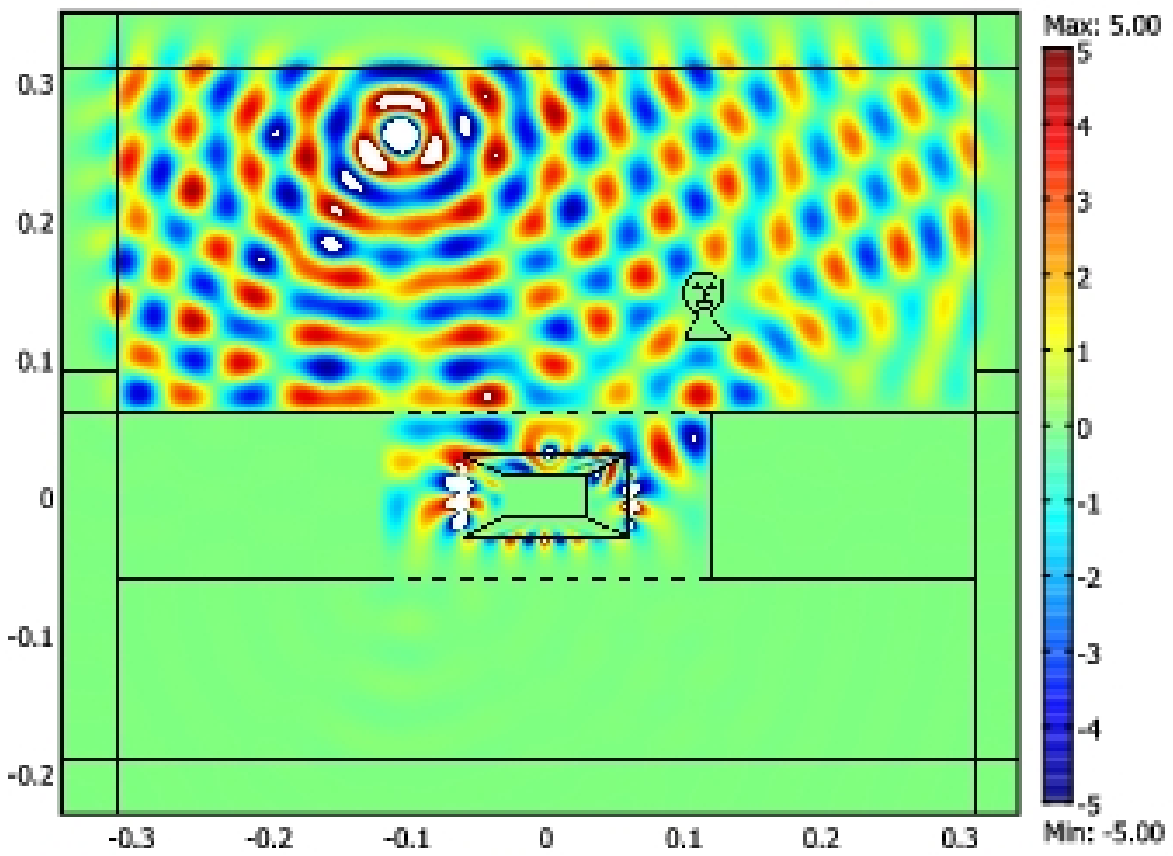}
\put(72,14){\bf(a)}
\end{overpic}
\begin{overpic}[scale=0.6,bb=0 0 352 264]{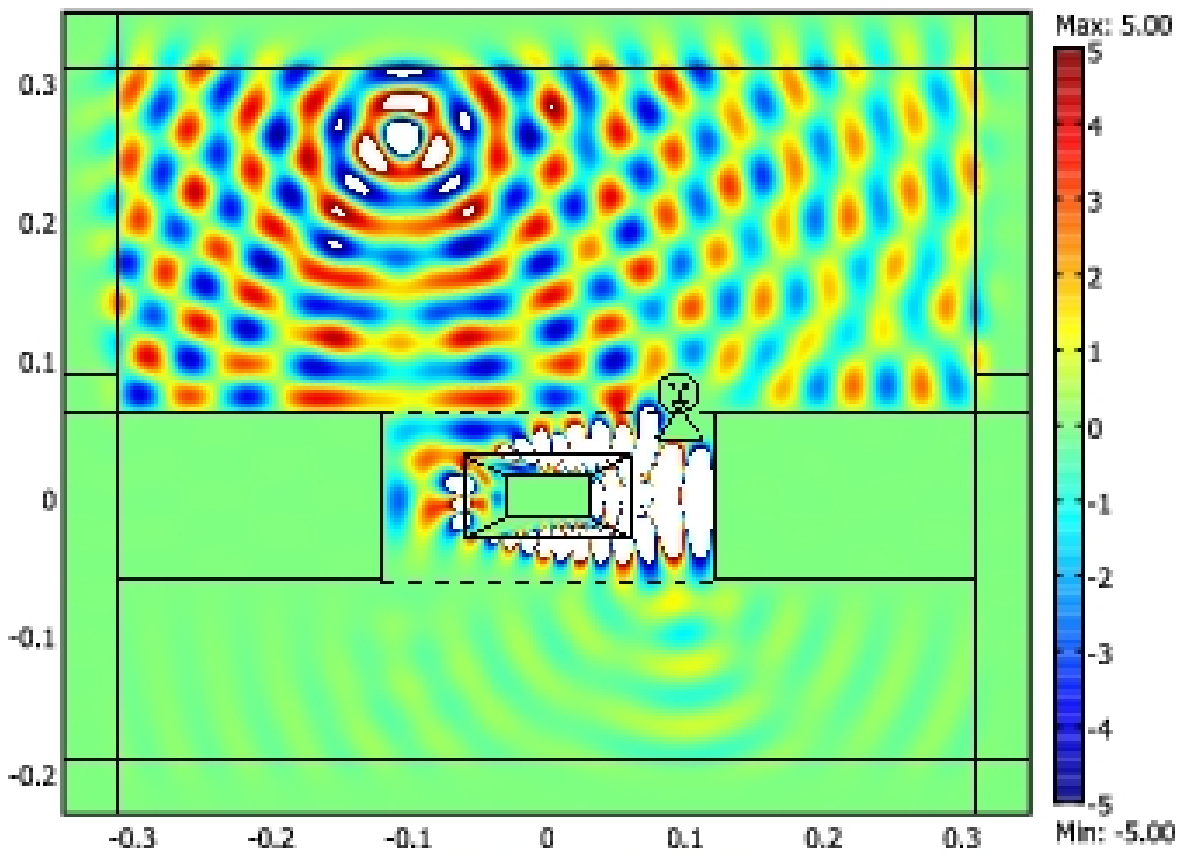}
\put(72,14){\bf(b)}
\end{overpic}
\begin{overpic}[scale=0.6,bb=0 0 352 264]{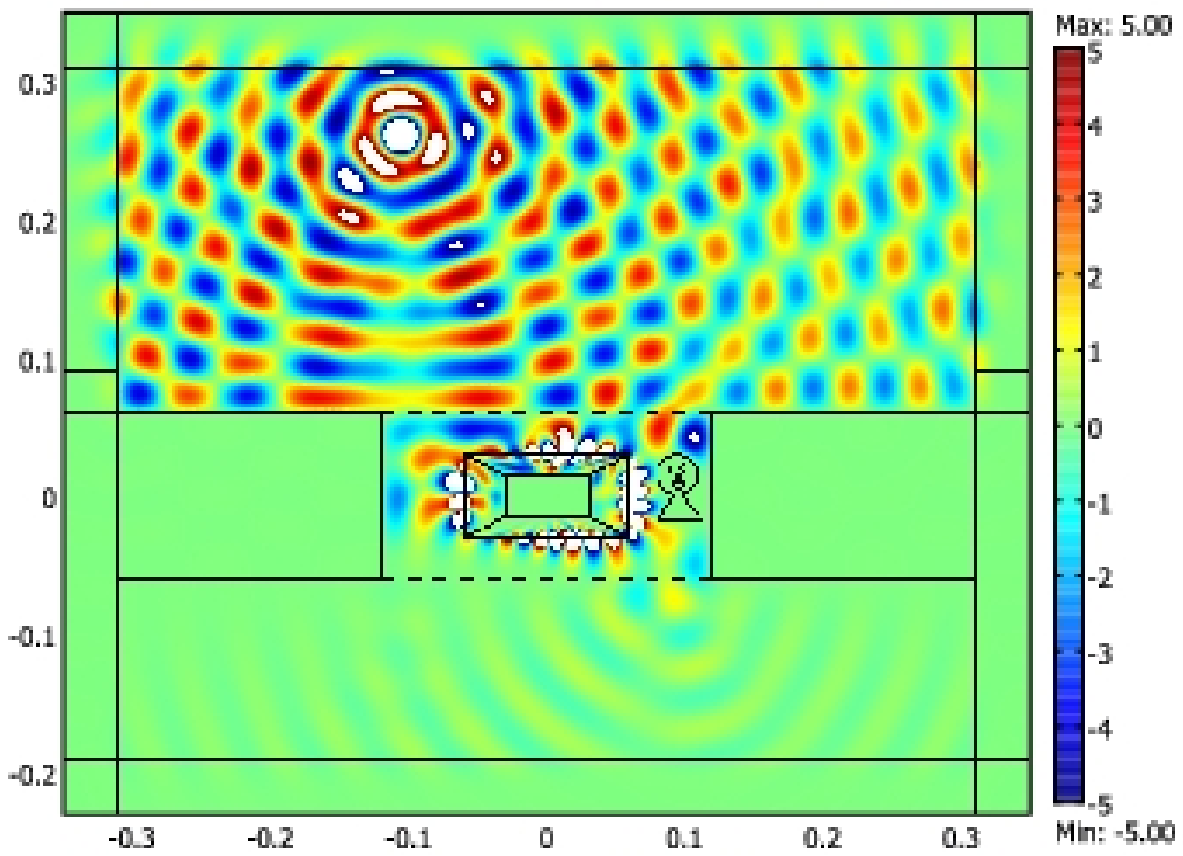}
\put(72,14){\bf(c)}
\end{overpic}
\begin{overpic}[scale=0.6,bb=0 0 352 264]{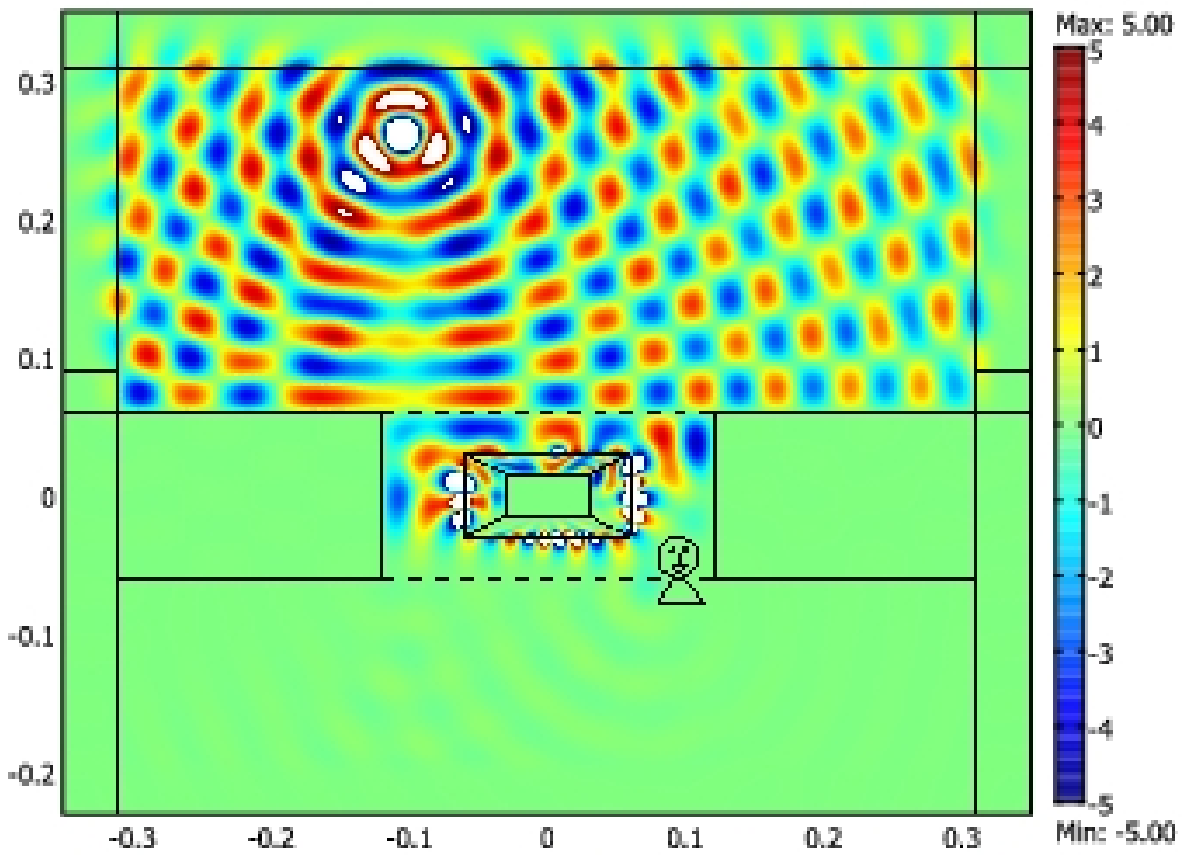}
\put(72,14){\bf(d)}
\end{overpic}
\caption{\label{steps} The total electric field distribution when
an object with PEC boundary go through the superscatterer
concealed entrance, the property of total reflection is broken
during the object pass through the entrance.}
\end{figure}

The concealed entrance is a successful case that rectangular
superscatterer can be regarded as building blocks, in which the
virtual rectangular cylinder is nicely fitted to the left and
right walls. More wider concealed entrance can be obtained if a
larger magnification factor $\eta$ is taken. Consequently, in the
region of microwave frequency, it is obvious that such a striking
device as ``platform nine and three-quarters'' at King's Cross
Station, where Harry Potter and his friends usually get on the
train to Hogwarts, is realizable. However, when another object is
placed in the virtual cylinder, the transformation optical
approach can not tell us what happens. In Fig. \ref{steps}, we
have simulated numerically the results when an object with PEC
boundary traverses the concealed entrance. Comparing the patterns
of electric field in the virtual cylinder and under the walls, we
find the concealment effect is destroyed more or less when object
is overlapped with the virtual cylinder.

In conclusion, we proposed a rectangular cylindrical
superscatterer and demonstrated its concealment effect on an
entrance in EM detection. The device performance are numerically
confirmed by full-wave simulations both in the far-field and
near-field. This kind functional devices can be regarded as
special building blocks, which are opaque by EM waves but
penetrable by small particles, and may be useful in concealing
important entrance.

The work is supported by the National Natural Science Foundation
of China under grant  No.10334020 and in part by the National
Minister of Education Program for Changjiang Scholars and
Innovative Research Team in University.

\end{document}